\documentclass[twocolumn,prl,superscriptaddress]{revtex4}
\AtBeginDocument{
\heavyrulewidth=.08em
\lightrulewidth=.05em
\cmidrulewidth=.03em
\belowrulesep=.65ex
\belowbottomsep=0pt
\aboverulesep=.4ex
\abovetopsep=0pt
\cmidrulesep=\doublerulesep
\cmidrulekern=.5em
\defaultaddspace=.5em
}

\usepackage{amsmath,amssymb}
\usepackage{graphicx}
\usepackage[usenames,dvipsnames]{xcolor}
\usepackage{amsthm}
\usepackage{booktabs}
\usepackage{hyperref}
\hypersetup{colorlinks=true,citecolor=Red,linkcolor=Blue,urlcolor=Black}
\usepackage{dsfont}
\usepackage{titlesec}

\begin{document}

\title {Computing Resonant Inelastic X-Ray Scattering Spectra Using The Density Matrix 
Renormalization Group Method}

\author{A. Nocera}
\affiliation{Department of Physics and Astronomy, The University of Tennessee, Knoxville, 
Tennessee 37996, USA}
\affiliation{Materials Science and Technology Division, Oak Ridge National Laboratory, Oak Ridge, Tennessee 37831, USA}

\author{U. Kumar}
\affiliation{Department of Physics and Astronomy, The University of Tennessee, Knoxville, 
Tennessee 37996, USA}
\affiliation{Joint Institute for Advanced Materials, The University of Tennessee, Knoxville, TN 37996, USA}

\author{N. Kaushal}
\affiliation{Department of Physics and Astronomy, The University of Tennessee, Knoxville, 
Tennessee 37996, USA}

\author{G. Alvarez}
\affiliation{Computational Science and Engineering Division and Center for Nanophase Materials Sciences, Oak Ridge National Laboratory, Oak Ridge, Tennessee 37831, USA}

\author{E. Dagotto}
\affiliation{Department of Physics and Astronomy, The University of Tennessee, Knoxville, 
Tennessee 37996, USA}
\affiliation{Materials Science and Technology Division, Oak Ridge National Laboratory, Oak Ridge, Tennessee 37831, USA}

\author{S. Johnston}
\affiliation{Department of Physics and Astronomy, The University of Tennessee, Knoxville, 
Tennessee 37996, USA}
\affiliation{Joint Institute for Advanced Materials, The University of Tennessee, Knoxville, TN 37996, USA}

\begin{abstract}
We present a method for computing resonant inelastic x-ray scattering (RIXS) spectra in one-dimensional systems using the density matrix renormalization group (DMRG) method.  By using DMRG to address the problem, we shift the computational bottleneck from the memory requirements associated with exact diagonalization (ED) calculations to the computational time associated with the DMRG algorithm. This approach is then used to obtain RIXS spectra on cluster sizes well beyond state-of-the-art ED techniques.  Using this new procedure, we compute the low-energy magnetic excitations observed in Cu $L$-edge RIXS for the challenging corner shared CuO$_4$ chains, both for large multi-orbital clusters and downfolded $t$-$J$ chains. We are able to directly compare results obtained from both models defined in clusters with identical momentum resolution.  In the strong coupling limit, we find that the downfolded $t$-$J$ model captures the features of the magnetic excitations probed by RIXS after a uniform scaling of the spectra is taken into account. 
\end{abstract}

\maketitle

Resonant inelastic x-ray scattering (RIXS) has emerged as a 
powerful and versatile probe of elementary excitations in quantum materials \cite{Ament2011,KotaniRMP2001}. 
One of the most commonly used approaches for computing RIXS spectra is 
small cluster exact diagonalization (ED) 
\cite{KotaniPRB2001,Kourtis2012,jia2014persistent,
MonneyPRL2013,JohnstonNatureComm2016,VernayPRB2008,
ChenPRL2010,SchlappaPreprint,Okada2006,Kuzian2012,Tsutsui2016,
Tohyama2015,Ishii2005,Tsutsui2003,Veenendaal1994,jia2012uncovering,FortePRB,KumarPreprint,WohlfeldPRB}.  
This approach is limited by the exponential growth of the Hilbert space, however, which 
restricts clusters to a relatively small size, thus limiting momentum resolution. 
For example, ED treatments of multi-orbital spin-chain systems such as 
the edge-shared CuGeO$_3$ or corner shared Sr$_2$CuO$_3$ have been limited to no 
more than six CuO$_4$ 
plaquettes \cite{VernayPRB2008,MonneyPRL2013,Kuzian2012,JJPRB2014,KotaniPRB2001}, 
while studies carried out using downfolded singleband Hubbard (or $t$-$J$) 
chains have been limited to $\sim 16 - 22$ sites \cite{SchlappaPreprint,Kourtis2012,
KumarPreprint,FortePRB}. 

\begin{figure}[ht]
\centering
\vskip -1cm
\includegraphics[width=\columnwidth]{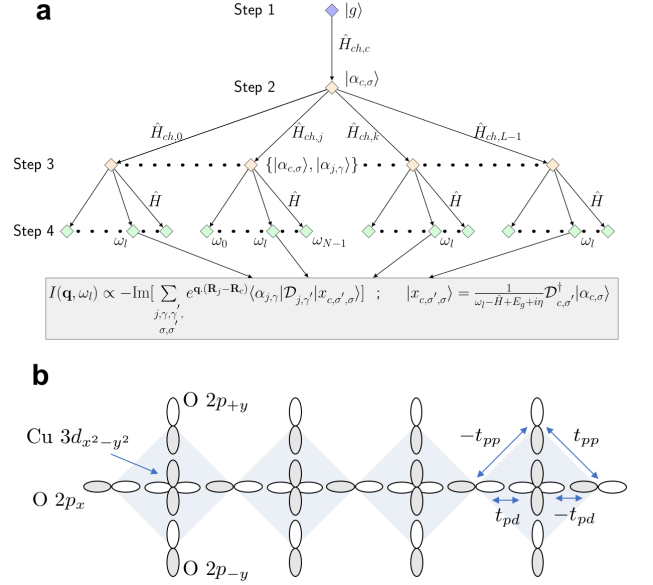}
\vskip -2cm
\caption{(a) A sketch of the algorithm for computing the real space 
Kramers-Heisenberg formula [Eq.~(\ref{RIXS3})] using the DMRG method at  
a fixed value of the energy loss $\Omega=\omega_l$. 
(b) A sketch of the multi-orbital $pd$-model describing the 
corner shared spin chain cuprates ({\it e.g.} Sr$_2$CuO$_3$).
}\label{algo_dmrg}
\end{figure}

The density matrix renormalization group (DMRG) is the most powerful 
method for computing the ground state properties of strongly 
correlated materials in one dimension (1D)~\cite{re:White1992,re:White1993,schollwock2011density}.
Within the DMRG framework, several efficient methods are available for computing dynamical correlation functions, 
including: time-dependent 
DMRG~\cite{re:Feiguin2004,re:Kollath2004}, which computes dynamical correlation 
functions in the time domain with a subsequent Fourier transform into frequency 
space~\cite{re:White2008}; correction-vector methods, which compute the dynamical 
correlator directly in frequency space~\cite{re:Kuhner1999,re:Jeckelmann2002,re:Jeckelmann2008,re:NoceraPRE2016}; 
continued fraction methods~\cite{re:Hallberg1995,re:Dargel2011,re:Dargel2012}; 
and Chebyshev polynomial expansion methods~\cite{re:Holzner2011,re:Wolf2015}. 
In this work, we present an efficient algorithm to compute the dynamical correlation 
function representing the RIXS scattering cross section with DMRG directly in frequency space. 
We then apply this approach to computing the Cu $L$-edge RIXS spectra of a quasi-1D 
corner-shared cuprate ({\it e.g.}, Sr$_2$CuO$_3$, see Fig. \ref{algo_dmrg}b), 
a geometry that is challenging for 
ED calculations due to significant finite size effects \cite{KotaniPRB2001,VernayPRB2008,MonneyPRL2013}. 
We consider a multi-orbital Hubbard model that retains the Cu and O orbital degrees of freedom, 
as well as a downfolded $t$-$J$ model. Using our DMRG-based approach, we access systems sizes beyond those accessible to ED, thus enabling us to directly compare the results obtained from the 
two models on large clusters with comparable momentum resolution.  

{\bf The Kramers-Heisenberg formalism} --- In a RIXS experiment, photons with 
energy $\omega_\mathrm{in}$ and momentum ${\bf k}_\mathrm{in}$ ($\hbar = 1$) scatter 
inelastically off of a sample, transferring momentum ${\bf q} = {\bf k}_\mathrm{out} - 
{\bf k}_\mathrm{in}$ and energy $\Omega = \omega_\mathrm{out}-\omega_\mathrm{in}$ to 
its elementary excitations. The resonant nature of the probe arises because 
$\omega_\mathrm{in}$ is tuned to match one of the elemental absorption edges,
such that it promotes a core electron to an unoccupied level of  
the crystal. 

The intensity of the RIXS process $I({\bf q},\Omega)$ is given by 
the Kramers-Heisenberg formalism~\cite{Ament2011,KotaniRMP2001}, with
\begin{equation}\label{RIXS1}
I({\bf q},\Omega) \propto |F_{f,g}|^2\delta(E_f - E_g + \Omega). 
\end{equation}
Here, $E_g$ and $E_f$ are the energies of the ground $|g\rangle$ and 
final states $|f\rangle$ of the system, respectively. 
The scattering amplitude $F_{f,g}$ is defined as
\begin{equation}\label{Eq:F}
F_{f,g} = \langle f|\hat{D}^{\dagger}(\textbf{k}_\mathrm{out})\frac{1}{\omega_\mathrm{in}-\hat{H}_\mathrm{ch}+E_g+\mathrm{i}\Gamma}\hat{D}(\textbf{k}_\mathrm{in})|g\rangle, 
\end{equation}
where $\hat{D}({\bf k})$ is the dipole transition operator describing the core-hole 
excitation. In what follows, we consider the Cu $L$-edge (a Cu $2p\rightarrow 3d$ transition). 
In this case, the dipole operator is defined as 
$\hat{D}_{j}(\textbf{k})=\sum_{\sigma,\alpha} e^{i \textbf{k}\cdot\textbf{R}_j} \left[A^{\hat{\epsilon}}_{\alpha} \hat{d}^{\dagger}_{j,\sigma} \hat{p}_{j,\alpha,\sigma}+\textrm{h.c.}\right]$,
where $d^{\dagger}_{j,\sigma}$ adds an electron to the valence band orbital ($3d_{x^2-y^2}$), and $\hat{p}_{j,\alpha,\sigma}$ 
destroys a spin $\sigma$ electron (creates a hole) in a core $2p_{\alpha}$ orbital on site $j$ 
located at $\textbf{R}_j$. The prefactor $A^{\hat{\epsilon}}_{\alpha}$ is the matrix
element of the dipole transition between the core $2p_{\alpha}$ orbital and the valence $3d_{x^2-y^2}$ orbital, $\langle 3d_{x^2-y^2,\sigma}|\hat{\epsilon}\cdot\hat{r}|2p_{\alpha,\sigma}\rangle$, which we set 
to $1$ for simplicity. $\Gamma$ is the inverse core-hole lifetime, 
and $\hat{H}_{\text{ch}}=\hat{H}+\hat{H}^C$, where  ${H}^C=V_C
\sum\limits_{j,\sigma,\sigma'} \hat{n}^d_{j,\sigma}
(1-\hat{n}^{p_{\alpha}}_{j,\sigma'})$ describes the Coulomb interaction
between the core hole and the valence electrons, and $\hat{H}$ is the many-body Hamiltonian 
of the system. 

Under the assumption that the core-hole is 
completely localized, and only one Cu $2p_{\alpha}$ orbitals is involved in the 
RIXS process, Eq. (\ref{Eq:F}) simplifies to 
\begin{equation}\nonumber
F_{f,g} \propto \sum\limits_{j,\sigma,\sigma'}e^{i \textbf{q}\cdot \textbf{R}_j}\langle f|\hat{D}^{\dagger}_{j,\sigma}\frac{1}{\omega_\mathrm{in}-\hat{H}_{\mathrm{ch},j}+E_g+i\Gamma} \hat{D}_{j,\sigma^{\prime}}|g\rangle, 
\end{equation}
where we have defined the local dipole-transition operator $\hat{D}_{j,\sigma}\equiv\hat{d}^\dagger_{j,\sigma}\hat{p}_{j,\sigma}$ and $\hat{H}_{\mathrm{ch},j}=\hat{H}+\hat{H}^C_j$, 
with $\hat{H}^{C}_j=V_C \sum\limits_{\sigma,\sigma'}\hat{n}^{d}_{j,\sigma}(1-\hat{n}^{p}_{j,\sigma'})$.

{\bf Reformulation of the problem for DMRG} --- The primary difficulty in
evaluating Eq. (\ref{RIXS1}) lies in computing the final states $|f\rangle$. 
This task is often accomplished using ED on small clusters meant to approximate the infinite
system. Obtaining these same final states is usually impossible with DMRG,
which \emph{targets} only the ground state; however, we will show that to
accomplish this task one can use the Lanczos method, which projects the state
onto a Krylov space~\cite{re:krylov31}.
Some of the present authors introduced this alternative method to calculate the
correction vectors for frequency-dependent correlation functions with DMRG
\cite{re:NoceraPRE2016}.  

We can formulate an efficient DMRG algorithm by expanding the square in Eq.~(\ref{RIXS1}), 
yielding a real space version of the Kramer-Heisenberg formula. 
To compact the notation, we define vectors  
$|\alpha_{j,\sigma}\rangle \equiv [\omega_\mathrm{in}-\hat{H}_{\mathrm{ch},j}+E_g+\mathrm{i}\Gamma]^{-1} \hat{D}_{j,\sigma}|g\rangle$.
Using this definition, Eq.~(\ref{RIXS1}) can be written as 
\begin{align}\label{RIXS3}
& I(\textbf{q},\Omega)\propto -\mathrm{Im} \Bigg[ \sum\limits_{i,j=0}^{L-1}\sum\limits_{\substack{\gamma,\gamma^{\prime}\\\sigma,\sigma^{\prime}}} e^{i \textbf{q}\cdot(\textbf{R}_i-\textbf{R}_j)}\times \nonumber \\
& \langle\alpha_{i,\gamma}| \hat{D}_{i,\gamma^{\prime}} \frac{1}{\Omega-\hat{H}+E_g+\mathrm{i}\eta} \hat{D}^{\dagger}_{j,\sigma^{\prime}}|\alpha_{j,\sigma}\rangle\Bigg].  
\end{align}
Here, $\eta$ is a broadening parameter, which plays the same role as the Gaussian or 
Lorentzian broadening introduced in ED treatments of the energy-conserving $\delta$-function 
appearing in Eq. (1). Throughout this work, we set it to $75$ meV.  
Note that the vectors  $|\alpha_{j,\sigma}\rangle$ 
must be computed for each value of $\omega_{\mathrm{in}}$ and $\Gamma$. 

The X-ray absorption spectrum (XAS) can be computed 
using a similar formalism. Its intensity is given by 
\begin{equation}\label{XAS}
I(\omega_\mathrm{in}) \propto -\mathrm{Im} \sum_{j,\sigma} \langle g | 
\hat{D}^{\dagger}_{j,\sigma}\frac{1}{\omega_\mathrm{in}-\hat{H}_{\mathrm{ch},j}+E_g+\mathrm{i}\Gamma}\hat{D}_{j,\sigma} |g\rangle.
\end{equation}

Finally, we note that we have removed the elastic line from all spectra shown in this work. The precise method for doing this is discussed in 
Supplementary Note IV. 

{\bf Computational procedure} --- The algorithm to compute the RIXS spectra 
using Eq. (\ref{RIXS3}) is as follows (see also Fig.~\ref{algo_dmrg}a): 

Step~1: Compute the ground state $|g\rangle$ of $\hat{H}$ using the standard ground state 
DMRG method. The vector $|g\rangle$ must be stored for later use. 

Step~2: \emph{Restart} from the ground state calculation, reading and then targeting the  
ground state vector calculated earlier and using a different 
Hamiltonian $\hat{H}_{\mathrm{ch},c}=\hat{H}+\hat{H}_{c}^{C}$, where $j = c$ is the center site of the 
chain. Construct the vector $|\alpha_{c,\sigma}\rangle$ at the \emph{center} 
of the chain using the Krylov-space correction vector approach~\cite{re:NoceraPRE2016} 
\begin{align}
|\alpha_{c,\sigma} \rangle &\simeq 
 \tilde{T}^{\dagger}_c \tilde{S}^{\dagger}_c \frac{1}{\omega_{\mathrm{in}}-D_{\mathrm{ch},c}+E_g+\mathrm{i}\Gamma}\tilde{S}_c \tilde{T}_c \hat{D}_{c,\sigma}|g\rangle, 
\end{align}
where we have performed a Lanczos tridiagonalization $\tilde{T}_c$ with starting 
vector $\hat{D}_{c,\sigma}|g\rangle$, and a subsequent diagonalization 
$\tilde{S}_c$ of the Hamiltonian $\hat{H}_{\mathrm{ch},c}$ and $D_{\mathrm{ch},c}$ 
is its diagonal form in the Krylov basis. 
The vector $|\alpha_{c,\sigma} \rangle$ should also be stored for later use.
Because the cluster is not periodic, the use of a central site here is an
approximation that will become exact in the thermodynamic limit. This central site ``trick'' 
was used for the first time in the application of time-dependent DMRG \cite{re:Feiguin2004}.  

Step~3: \emph{Restart} from previous run, now using a different Hamiltonian $\hat{H}_{\mathrm{ch},j}=\hat{H}+\hat{H}_{j}^{C}$. 
Read and then target (in the DMRG sense) the ground state vector 
calculated in Step~1, as well as the vector $|\alpha_{c,\sigma}\rangle$ constructed 
in Step~2. For each site $j$, except for the center site considered in Step~2, construct the 
vector 
\begin{align}
|\alpha_{j,\gamma}\rangle &\simeq  
  \tilde{T}^{\dagger}_j\tilde{S}^{\dagger}_j\frac{1}{\omega_{\mathrm{in}}-D_{\mathrm{ch},j}+E_g+\mathrm{i}\Gamma}\tilde{S}_j \tilde{T}_j \hat{D}_{j,\gamma}|g\rangle,  
\end{align}
with Lanczos tridiagonalization $\tilde{T}_j$ with starting vector $\hat{D}_{j,\gamma}|g\rangle$, and a subsequent 
diagonalization of $\hat{H}_{\text{ch},j}$. This step of the algorithm requires 
a number of runs which is equal to the number of sites
minus $1$, {\it i.e.}, $L-1$. These can be run in parallel on a standard cluster machine, 
restarting from Step~2. Performing Step 2 and Step 3 in this sequence is
crucial for having the vectros $|\alpha_{c,\sigma}\rangle$ and
$|\alpha_{j,\gamma}\rangle$ in the \emph{same} DMRG basis. 
The vector $|\alpha_{j,\gamma} \rangle$ should also be stored for later use.

Step~4: \emph{Restart} using the original 
Hamiltonian $\hat{H}$. Read and then target the ground state produced in 
Step~1, $|\alpha_{c,\sigma} \rangle$ produced in Step~2, and the vector
$|\alpha_{j,\gamma}\rangle$ constructed in Step~3. 
For a fixed $\Omega=\omega_l$, compute the correction vector of $|\alpha_{c,\sigma}\rangle$ using again the Krylov-space correction vector 
approach as 
\begin{align}
| & x_{c, \sigma^{\prime},\sigma} \rangle \equiv \frac{1}{\Omega-\hat{H}+E_g+\mathrm{i}\eta} \hat{D}^{\dagger}_{c,\sigma^{\prime}}|\alpha_{c,\sigma}\rangle \nonumber\\
&= \tilde{T}^{\dagger}\tilde{S}^{\dagger}\frac{1}{\Omega-D+E_g+\mathrm{i}\eta}\tilde{S} \tilde{T} \hat{D}^{\dagger}_{c,\sigma^{\prime}}|\alpha_{c,\sigma}\rangle 
\end{align}
with Lanczos tridiagonalization $\tilde{T}$ (using
$\hat{D}^{\dagger}_{j,\sigma^{\prime}}|\alpha_{c,\sigma}\rangle$ as the seed) 
and a subsequent diagonalization $\tilde{S}$ of the Hamiltonian $\hat{H}$, with $D$ being the 
diagonal form of $\hat{H}$ in the Krylov basis.
This is a crucial part of the algorithm, which amounts to 
computing the correction vector $|x_{c, \sigma^{\prime},\sigma}\rangle$
of a previously calculated correction vector $|\alpha_{c,\sigma}\rangle$.
Execute this computation $N_{\Omega}$ times for $\Omega\in[\omega_0, \omega_{N-1}]$. 

Step~5: Finally, compute the RIXS spectrum in real space 
$I_{j,c}(\Omega)\propto \langle \alpha_{j,\gamma}| \hat{D}_{j,\gamma^{\prime}} |x_{c,\sigma^{\prime},\sigma}\rangle $ 
and then Fourier transform the imaginary part to obtain the RIXS intensity
\begin{equation}\label{RIXScenter}
I(\textbf{q},\Omega)\propto -\mathrm{Im} \sum\limits_{\substack{j,\gamma,\gamma^{\prime}\\\sigma,\sigma^{\prime}}}e^{i \textbf{q}\cdot(\textbf{R}_j-\textbf{R}_c)} I_{j,c}(\Omega).
\end{equation}

\begin{figure}[thbp] 
\centering
\includegraphics[width=8.75cm]{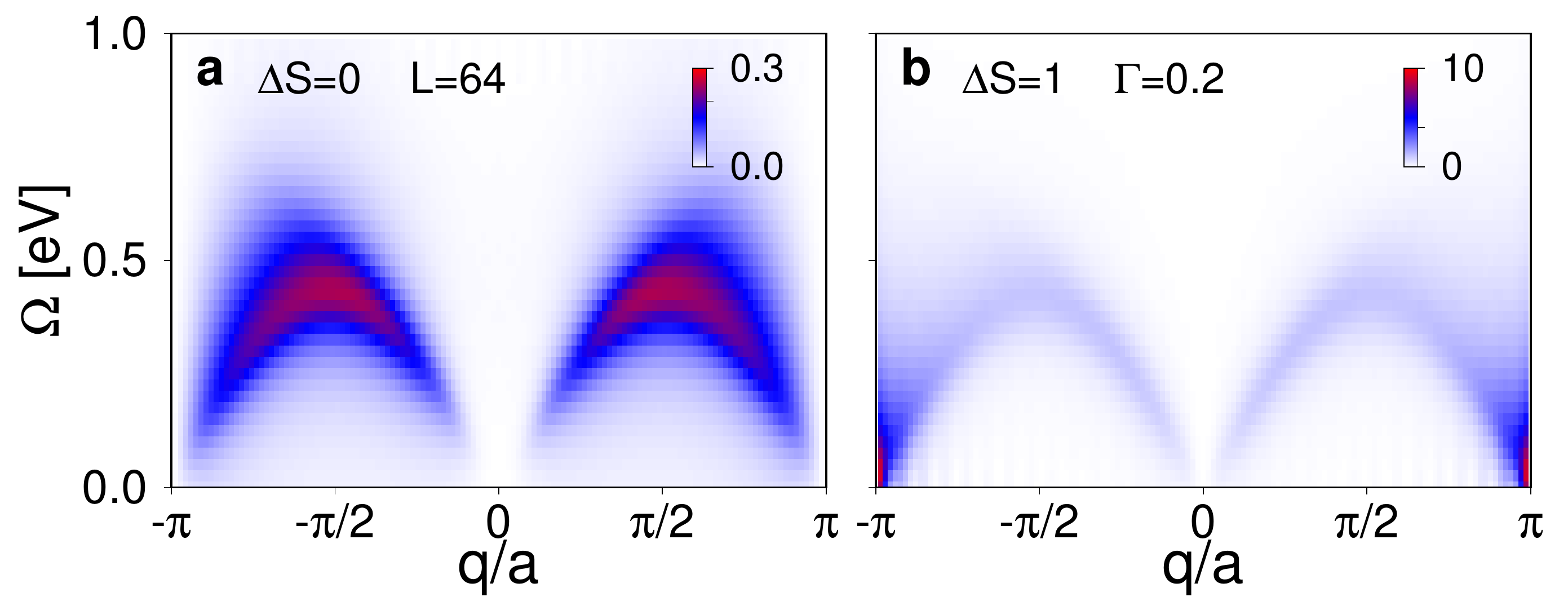}
\caption{DMRG results for the RIXS intensity $I(q,\Omega)$ of a half-filled $t$-$J$ chain. 
Results are shown for a $L=64$ site 
chain, in the (a) $\Delta S = 0$ and (b) $\Delta S = 1$ channels. 
The remaining parameters are $t = 0.3$~eV, $J=0.25$~eV, $\eta = 75$~meV
and $\omega_{\mathrm{in}}=0.1$~eV (which corresponds to the resonance observed in the 
XAS).}\label{fig1}
\end{figure}

{\bf Computational complexity} --- The computational cost required for DMRG to compute the
RIXS spectrum can be easily estimated, assuming that the ground state of the Hamiltonian has already been calculated. 
Let $C_{2-3}$ be the 
computational cost ({\it i.e.,} the number of hours) for a single run in
Step~2 ($1$ run only) or Step~3 ($L-1$ runs in total). 
Let $C_{4}$ be the computational cost for a single run in Step~4.
The total computational time needed to compute 
the RIXS spectrum is then 
$\textrm{CPU}_\text{cost}=C_{2-3}L + C_{4} L N_{\Omega}$, where $N_{\Omega}$ 
is the number of frequencies needed in a given interval of energy losses. 
The use of this center site ``trick" reduces the computational cost by a factor 
of the order of $L$ (Eq.~(\ref{RIXS3}) to Eq.~(\ref{RIXScenter})).
For the largest system size considered in this work ($20$ plaquettes in CuO$_4$ multi-orbital model 
at half-filling, using up to $m=1000$ DMRG states), 
the typical values for $\textrm{CPU}_\text{cost}$ on a single core of a standard computer 
cluster are: $C_{2-3}\sim 2$~hours, while $C_{4}\sim 2-24$~hours. 
The computational cost $C_4$ for Step~4 follows the typical performance 
profile of the Krylov-space approach found in Ref.~\cite{re:NoceraPRE2016}, where less CPU time is needed to compute the spectra at lower energy-losses. We also note that the calculation of each energy loss is trivially parallelizable. 
From these assumptions, we estimate the proposed method can compute the RIXS
spectrum of a cluster as large as Cu$_{20}$O$_{61}$ in less than a day if enough cores are available. 

{\bf Numerical Results for the $t$-$J$ model} --- 
We first apply our approach to compute the RIXS spectrum of the 1D $t$-$J$ model 
as an effective model for the antiferromagnetic corner-shared spin chain 
cuprate Sr$_2$CuO$_3$ (see Methods). Throughout this paper, we adopt open boundary conditions and 
work at half-filling and set $t=0.3$~eV for the nearest neighbor hopping 
and $J=0.25$~eV for the antiferromagnetic exchange interaction. These values are typical for 
Sr$_2$CuO$_3$~\cite{Suzuura96,Motoyama96,Kojima97,schlappa2012spin,walters2009effect,Lee2013,Bisogni2012}. 

Before scaling up our DMRG calculations to large systems, we benchmarked our method by directly comparing our DMRG results to ED. 
The results for a $L = 16$ sites $t$-$J$ chain are presented in Supplementary Note I. 
(We provide a similar comparison for a four-plaquette 
multi-orbital cluster in Supplementary Note II.)
Our DMRG approach gives perfect agreement with the ED result for both the 
XAS and RIXS spectra, for the largest clusters we can access with ED.  
All of the DRMG simulations presented in this work used up to $m=1000$ states, 
with a truncation error smaller than $10^{-6}$. 

\begin{figure*}[thbp]
\centering
\vskip -9.5cm
\includegraphics[width=18cm]{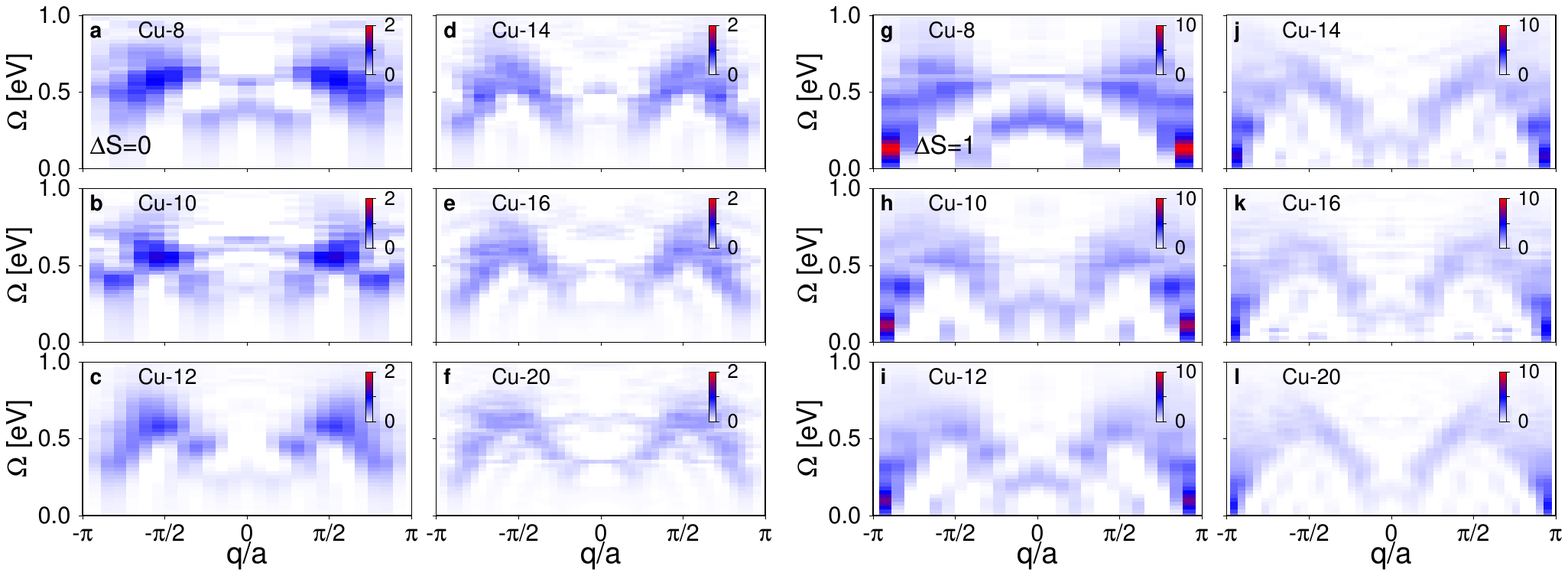}
\vskip -9.5cm
\caption{ The response $I(q,\Omega)$ obtained with DMRG for a multi-orbital \emph{pd} model as a 
function of the number of CuO$_4$ plaquettes. Panels 
(a)-(f) show the spin-conserving $\Delta S=0$ channel to the RIXS 
intensity, while panels (g)-(l) show the non-spin-conserving $\Delta S = 1$  
channel. Results are shown for 8 to 20 unit cells at half filling, 
computed at resonance with $\omega_{\mathrm{in}}=2.5$, $\Gamma=0.2$, and $V_C=4.0$.
}\label{fig2}
\end{figure*}

We now turn to results obtained on a $L=64$ site chain, 
as shown in Fig.~\ref{fig1}. Here, we present results for the spin-flip ($\Delta S=1$) 
and non-spin-flip ($\Delta S=0$) contributions to the total RIXS intensity. 
The $\Delta S=0$ contribution corresponds to the $\sigma=\sigma^\prime$ and 
$\gamma=\gamma^\prime$ terms in the Kramers-Heisenberg formula Eq.~(\ref{RIXS3}). 
In this case, only two configurations 
($\gamma=\gamma^{\prime}=\sigma=\sigma^{\prime}=\uparrow$ and 
$\gamma=\gamma^{\prime}=\downarrow$, $\sigma=\sigma^{\prime}=\uparrow$) have to be explictly 
calculated with DMRG, as the other two possible spin conserving configurations contribute equally by 
symmetry. 
The remaining terms with $\sigma\neq\sigma^\prime$ and $\gamma\neq\gamma^\prime$ determine 
the non-spin conserving $\Delta S=1$ contributions to the spectrum. 
In this case, only one configuration ($\sigma^{\prime}=\downarrow$, 
$\sigma=\uparrow$, $\gamma^{\prime}=\downarrow$, $\gamma=\uparrow$) has been
simulated with DMRG, as the \emph{flipped} 
configuration ($\sigma^{\prime}=\uparrow$, $\sigma=\downarrow$,
$\gamma^{\prime}=\uparrow$, $\gamma=\downarrow$) contributes equally by
symmetry. The remaining two possible non-spin conserving configurations also 
give zero contribution to the RIXS spectrum by symmetry. 

In Fig.~\ref{fig1}, the $\Delta S=1$ part of the RIXS spectrum shows a continuum 
of excitations resembling the two spinon continuum commonly observed in the dynamical spin 
structure factor $S(q,\omega)$ of one-dimensional spin-$1/2$ 
antiferromagnets~\cite{Tennant95,lake2005quantum,mourigal2013fractional,Lake2013}. 
The $\Delta S=0$ contribution in Fig \ref{fig1}a shows two broad arcs with maxima at $q = \pi/2a$. 
Notice also a perfect cancellation of the RIXS signal at the zone boundary, 
which is $q=\frac{\pi}{a}\frac{L}{L+1}$ in open boundary conditions.  
Our results agree with the ED results of Refs. \onlinecite{Kourtis2012} and \onlinecite{FortePRB}, 
but with much better momentum resolution. 
We find that the finite size effects of the magnetic excitations in the $t$-$J$ 
model are mild; we observe only small differences between results obtained on $L=32$  (not shown in Fig.~\ref{fig1}) and $L=64$ site clusters.  

{\bf Magnetic excitations in the multi-orbital $pd$-model} --- 
In the strong coupling limit, the low-energy magnetic response of the spin-chain 
cuprates are believed to be effectively described by a single orbital 
Hubbard or $t$-$J$ model~\cite{Emery87,ZhangRice88}. 
According to this picture, \emph{holes} predominantly occupy the Cu
orbitals at half-filling, while the oxygens along the Cu-Cu direction provide a
pathway for superexchange interactions between the nearest-neighbor Cu
orbitals. Since our DMRG approach provides access to large cluster sizes, 
we now compute the RIXS spectrum of a more realistic multi-orbital model. 
Here, we consider the challenging corner-shared geometry, which suffers from slow convergence in the cluster size. 
To address this, we consider finite 1D Cu$_n$O$_{3n+1}$ clusters, with open boundary
conditions, as illustrated in Fig.~\ref{algo_dmrg}b for the $n = 4$ case. 
The Hamiltonian is given in the Methods section.  
We evaluated the Cu $L$-edge RIXS intensity for this model as a function of $n$ for 
up to $n = 20$ CuO$_4$ plaquettes. 

The RIXS spectra for spin-conserving ($\Delta S=0$) and
non-spin-conserving contributions ($\Delta S=1$) calculated with our DMRG
method are shown in Fig.~\ref{fig2}. 
Similar to the $t-J$ spectra, panels (a-f) in Fig.~\ref{fig2} show two 
broad arcs with maxima at $\pm \pi/2a$.
Here, we observe significant finite size effects in the RIXS spectra. Some of
these effects are the result of our use of the ``center-site approximation" in
evaluating the Kramers-Heisenberg formula. For example, the downward
dispersing low-energy peak centered at $q = 0$ seen in the smaller clusters is
the result of this approximation. These features in the spectra can be
minimized by carrying out calculations on larger clusters. 
Because of this, to observe well defined spectral features, we need to consider
at least fourteen plaquettes. The \emph{pd} model also shows that the
low-energy $\Delta S=1$ part of the RIXS spectrum is characterized by a
two-spinon-like continuum of excitations (panels (g-l) in Fig.~\ref{fig2}).

{\bf Comparing the multi-orbital and effective $\boldmath{t}$-$\boldmath{J}$ models} ---  
Over the past decade, there has been a considerable research effort dedicated
to quantitatively understand the intensity of magnetic excitations probed by
inelastic neutron scattering (INS) \cite{NaglerPRB,walters2009effect,mourigal2013fractional}. 
This effort is motivated by the desire to understand the relationship between the spectral weight of the
dynamical spin response $S({\bf q},\omega)$ and the superconducting transition
temperature T$_c$ of unconventional superconductors \cite{DougRMP}. To this end, several
studies have set out to determine whether the observed INS intensity can be
accounted for by the Heisenberg model in low-dimensional strongly correlated
cuprates. Here, the highest degree of success has been achieved in quasi-1D
materials, where accurate theoretical predictions for $S(q, \omega)$ are
available \cite{walters2009effect,mourigal2013fractional}. Many of these studies find that the low-energy Heisenberg
model can indeed account for the INS intensity, after accounting for
corrections due to effects such as the degree of covalency, its impact on the form
factor, and Debye-Waller factors. 

RIXS has also been applied to study magnetic excitations in many of the same
materials \cite{schlappa2012spin,SchlappaPreprint,Bisogni2012}. It is therefore natural to ponder how covalency
modifies the magnetic excitations as viewed by RIXS. In the limit of a short
core-hole lifetime, or under constraints in the incoming and outgoing photon
polarization, the RIXS intensity for single orbital Hubbard and $t$-$J$ chains 
is well approximated by $S(q, \omega)$ 
\cite{jia2014persistent,FortePRB,Bisogni2012,Ament2011}.
However, to the best of our knowledge, no systematic comparison of the RIXS
intensity, as computed by the Kramers-Heisenberg formalism, has been carried
out for multi-orbital and downfolded Hamiltonians. 

\begin{figure}[thbp]
\centering
\vskip -1cm
\includegraphics[width=8.5cm]{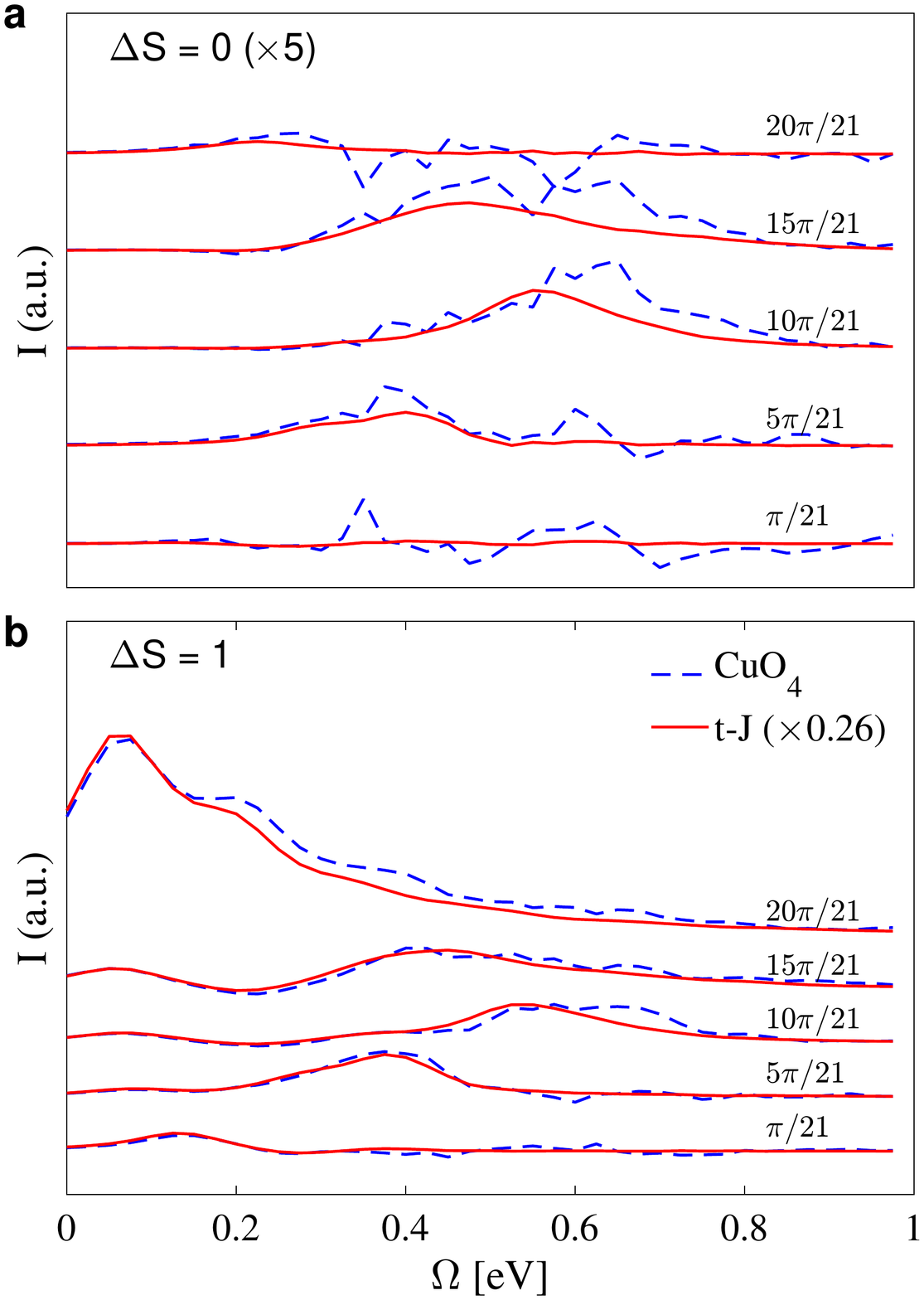}
\vskip -1.5cm
\caption{
A comparison of the magnetic RIXS excitations computed using DMRG 
for a 20-site $t$-$J$ chain (solid red line) and multi-orbital \emph{pd} model (dashed blue line) with
twenty unit cells at half filling. Results are shown for the (a) $\Delta S=0$
and (b) $\Delta S=1$ channels. 
The parameters for the $t$-$J$ model are $t=0.5$~eV, $J=0.325$~eV, $V_C = 4$~eV, $\Gamma = 0.2$~eV,  
and $\omega_{\mathrm{in}}=0.14$~eV. 
The parameters for the multi-orbital model are given in the main text. 
The incident photon energy is $\omega_{\mathrm{in}}=2.5$~eV, the inverse core hole lifetime is $\Gamma=0.2$~eV, 
and the core hole potential is $V_C=4$~eV. 
The results for the $t$-$J$ model have been scaled by a factor of $0.26$ such that the maximum 
intensity of the $\Delta S = 1$ excitations are the same at the zone boundary. 
} \label{fig3}
\end{figure}

Figure \ref{fig2} demonstrates that DMRG grants access to large system sizes. 
We are, therefore, in a position to make such a comparison for the multi-orbital spin-chain cuprates. Figure~\ref{fig3} 
compares the spectra computed on a $L = 20$ site $t$-$J$ chain against those computed on a Cu$_{20}$O$_{61}$ cluster, such that the momentum resolution of the two clusters is the same. 
The parameters for the multi-orbital model are identical to those used in 
Fig.~\ref{fig2}. To facilitate a meaningful comparison with the $t$-$J$ model, 
we adopted $t=0.5$~eV and $J = 0.325$~eV. These values are obtained by diagonalizing a 
Cu$_2$O$_7$ cluster (see methods). Note that we use the same value of the core hole potential 
$V_C = 4$~eV in both cases. In supplementary note III, we show results for a reduced value of 
$V_C$ for the $t$-$J$ model, which are very similar.   
To compare the two spectra, the results for the $t$-$J$ model have been scaled by a factor of $0.26$ such that the maximum intensity of the $\Delta S = 1$ excitations is the same at the zone boundary. This factor presumably accounts for covalent factors and differences in how the core-hole interacts with the distribution of electrons in the intermediate state. 

After we have rescaled the spectra, we find excellent overall agreement between the two 
calculations: the amplitude of the broad arcs for the magnetic excitations, both in the $\Delta
S=0$ and in $\Delta S=1$ channels of the RIXS spectra are well captured by the
$t$-$J$ model. There are, however, minor quantitative differences related to the
spectral weight of the excitations appearing near $q = \pi/2a$ in the $\Delta S =
0$ channel.  
For example, the $t$-$J$ model concentrates the magnetic
excitations at slightly lower values of the energy loss in $\Delta S = 0$ channel. 
This discrepancy might be compensated for by taking a different value of $J$; however, this would come at the expense of 
the agreement in the $\Delta S = 1$ channel. 
These differences should be
kept in mind when one calculates the low-energy magnetic RIXS spectra using an
effective $t$-$J$ or single-band Hubbard model. Nevertheless, our results show that in the strong coupling limit, the magnetic RIXS spectrum can be described well by the effective $t$-$J$ model.  

Figure \ref{fig3} shows that that the overall agreement between 
the full  multi-orbital model and the $t$-$J$ model is much better in the 
$\Delta S = 1$ channel than in the $\Delta S = 0$ channel. 
We can naively understand this difference by recalling the role of charge
fluctuations in the two magnetic excitation pathways. The $\Delta S = 1$ RIXS
excitations are possible in a system with strong spin-orbit coupling in the Cu
$2p$ orbitals, which allows the spin of the core-hole to flip in the
intermediate state of the RIXS process
\cite{FortePRB,SchlappaPreprint,Kourtis2012}. The $\Delta S = 0$ pathway,
however, requires a double spin-flip between neighboring Cu spins in the final state 
\cite{FortePRB,Kourtis2012}. At the Cu $L$-edge, 
such processes occur due to charge fluctuations between the 
neighboring Cu sites in the intermediate state. The multi-orbital model treats such 
charge fluctuations differently owing to the presence of the ligand oxygen orbitals. 
This difference accounts for the discrepancy between the two models in the $\Delta S = 0$ channel.   
At the Cu $L$-edge, however, the strong core-hole potential suppresses this difference by repelling holes from the site where it was created resulting in only minor differences between the predictions of the two models. 

{\bf Concluding Remarks} --- We have presented a novel 
DMRG approach to computing the RIXS spectra and benchmarked this method against traditional ED. Using our DMRG algorithm, we can compute the 
RIXS spectra on 1D clusters much larger than those accessible to state-of-the-art ED methods. 
Using this method, we modeled the magnetic excitations probed by RIXS at the Cu $L$-edge in 
1D antiferromagnets on the largest cluster sizes to date. 
We found that both the full multi-orbital cluster and the effective $t$-$J$ model provide comparable descriptions of the excitations in the $\Delta S = 1$ channel, while there were minor quantitative differences in the $\Delta S = 0$ channel.
These differences could be explained by noting the difference in the way that
these two channels probe magnetic
excitations. 
Finally, we note that the bottleneck to RIXS simulations using ED is the exponential growth 
of the Hilbert space. Our approach shifts the computational burden to the availability of 
CPUs thus opening the door to calculations for large systems.  
For example, one can envision extending this approach to the quasi-2D
models currently under active study by the DMRG community. 

{\bf Methods} --- 
The multi-orbital $pd$-Hamiltonian describing the corner-shared spin-chains, given in the hole-picture, is
\begin{equation}\label{Hamipd_part1}
\begin{aligned}
H &=\epsilon_d\sum\limits_{i,\sigma}  n^d_{i,\sigma}+\sum_{j,\sigma}  \epsilon_{p,\gamma}n^p_{j,\gamma,\sigma} 
   +\sum\limits_{\substack{\langle i,j\rangle\\ \gamma,\sigma}} t_{pd}^{ij} (d^\dagger_{i,\sigma} p_{j,\gamma,\sigma} + \textrm{h.c.})\\
  &+\sum\limits_{\substack{\langle j,j'\rangle\\\gamma,\gamma^\prime,\sigma}} t_{pp}^{jj^\prime} p^\dagger_{j,\gamma,\sigma}p_{j^\prime,\gamma^\prime\sigma}
   +U_{d} \sum\limits_i n^d_{i,\uparrow} n^d_{i,\downarrow} \\
 &+ U_{p}\sum\limits_{i,\gamma}  n^p_{j,\gamma,\uparrow} n^p_{j,\gamma,\downarrow}
  + U_{pd}\sum\limits_{\substack{\langle i,j\rangle \\ \sigma,\sigma^\prime}}
   n^d_{i,\sigma} n^p_{j,\gamma,\sigma^\prime}.
\end{aligned}
\end{equation}
Here, $\langle\dots\rangle$ denotes a sum over nearest neighbor orbitals;
$d^{\dagger}_{i,\sigma}$ ($p^{\dagger}_{j,\gamma,\sigma}$) creates a spin $\sigma$ hole
on the $i^\mathrm{th}$ Cu 3$d_{x^2-y^2}$ orbital (the $j^\mathrm{th}$ O 2$p_{\gamma}$ orbital, 
$\gamma = x$,~$\pm y$); 
$\epsilon_d$ and $\epsilon_p$ are the on-site energies; $n^d_{i,\sigma}$
($n^p_{j,\gamma,\sigma}$) is the number operator for
the Cu 3$d_{x^2-y^2}$ orbital (the $j^\mathrm{th}$ O 2$p_{\gamma}$ orbital);
$t_{pd}^{ij}$ and $t_{pp}^{j,j^\prime}$ are the Cu-O and O-O overlap integrals,
respectively; $U_d$ and $U_p$ are the onsite Hubbard repulsions of the Cu and O
orbitals,
respectively, and $U_{pd}$ is the nearest-neighbor Cu-O Hubbard repulsion. 
The phase convention for the overlap integrals is shown in Fig. \ref{algo_dmrg}b. 
In this work, we adopt (in units of eV) $\epsilon_d = 0$,
$\epsilon_{p,x} = 3$, $\epsilon_{p,y} = 3.5$, 
$|t_{(p,x)d}| = 1.5$ $|t_{(p,y)d}| = 1.8$, $|t_{pp}| = 0.75$,
$U_d = 8$, $U_p = 4$, and $U_{pd} = 1$, following Ref. \onlinecite{WohlfeldPRB}.  

In the limit of large $U_d$, one integrates out the oxygen degrees of freedom and 
maps Eq. (\ref{Hamipd_part1}) onto an effective spin-$1/2$ $t$-$J$ Hamiltonian \cite{ZhangRice88} 
\begin{equation*}
 H = -t \sum_{ i , \sigma}(\tilde{d}_{ i, \sigma }^\dagger 
 \tilde{d}_{i+1,\sigma}^{\phantom\dagger}+h.c.) 
 + J\sum_{ i }\textbf{S}_i\cdot \textbf{S}_{i+1}.
 \end{equation*}
Here, $\tilde{d}_{i,\sigma }$ is the annihilation operator for a hole with spin
$\sigma$ at site $i$, under the constraint of no double occupancy, $n_i = 
\sum_\sigma n_{i,\sigma}$ is the number operator, and $\textbf{S}_i$ is the
spin operator at site $i$. 

To facilitate a direct comparison between the two models, 
one can extract the hopping $t$ and exchange interaction $J$  
from an ED calculation of a two-plaquette Cu$_2$O$_7$ cluster with open boundary conditions \cite{JohnstonEPL}. 
Here, we obtain the hopping ($t=0.5$~eV) by diagonalizing cluster in the 
($2\uparrow,1\downarrow$)-hole sector, and setting $2t$ to be equal to the  
energy separation between the bonding and antibonding states of the 
Zhang-Rice singlet. Similarly, we can obtain the 
superexchange ($J=0.325$~eV) by diagonalizing the cluster in the 
$(1\uparrow,~1\downarrow)$-hole sector, and setting  
the singlet-triplet splitting of the Cu ($d^9d^9$) 
configurations equal to $J$.  

{\bf Acknowledgements} --- A. N. and E. D. were supported by the U.S. Department of Energy, Office of Basic Energy Sciences, Materials Sciences and Engineering Division. G. A. and S. J. were supported by the Scientific Discovery through Advanced Computing (SciDAC) program funded by the U.S. Department of Energy, Office of Sciences, Advanced Scientific Computing Research and Basic Energy Sciences, Division of Materials Sciences and Engineering. This  research  used  computational resources supported both  
by the University of Tennessee and Oak Ridge National Laboratory Joint Institute for Computational Sciences (Advanced Computing Facility). It also used computational resources at the National Energy Research Scientific Computing Center (NERSC).

\onecolumngrid
\newpage
\setcounter{figure}{0}
\setcounter{equation}{0}
\titleformat{\section}{\normalfont\Large\bfseries}{Supplementary Note~\thesection:}{1em}{}
\renewcommand{\figurename}{Supplementary Figure }

\section{Supplementary Note~I:~Benchmarks on a 16-site $t$-$J$ chain}
\begin{figure}[thbp]
\centering
\vskip -4.0cm
\includegraphics[width=0.7\textwidth]{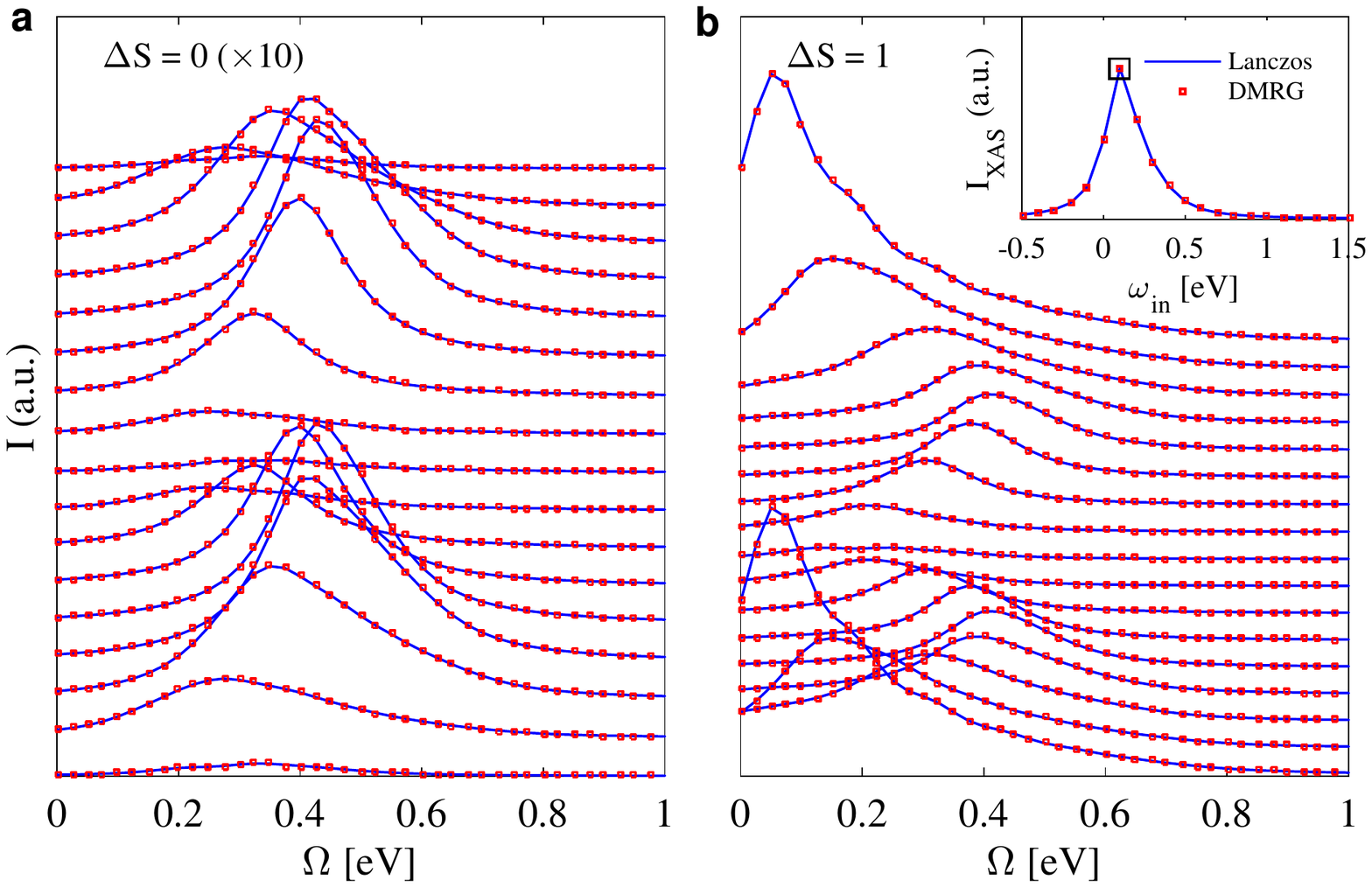}
\vskip -4.5cm
\caption{A comparison of the RIXS response $I(q,\Omega)$ 
calculated for a $t$-$J$ chain of length $L=16$ at half filling $N=16$ 
with $J = 0.25$~eV and $t=0.3$~eV.  
Results are shown for both the (a) $\Delta S = 0$ and (b) $\Delta S = 1$ 
channels, computed at resonance with $\omega_{\mathrm{in}}=0.1$~eV. 
We have computed the spectrum with DMRG (red squares), which we compare to results obtained from Lanczos ED (solid blue line). 
The inset of panel (b) compares the XAS spectrum 
computed with DMRG [using Eq. (8)] and Lanczos ED. The 
open black box indicates the incident energy $\omega_{\mathrm{in}}$ 
used to compute the RIXS spectra. Note that in this comparison we did not use the 
center trick for calculating the RIXS spectra. Instead, Eq.~(3) of the main text has been used.  
} \label{supp_fig1}
\end{figure}

In this note, we compare the results of our DMRG method against the spectrum obtained from Lanczos ED. 
Supplementary Figure~\ref{supp_fig1} directly compares the results from the two
methods applied to a $L = 16$ site $t$-$J$ chain, where our DMRG approach gives
perfect agreement with the ED results for both the XAS and RIXS spectra. Here, 
we have assumed parameter values typical for a Cu $L$-edge 
measurement performed on Sr$_2$CuO$_3$ with $t=0.3$, $J=0.25$,  
an inverse core hole lifetime $\Gamma=0.3$~eV, and a core-hole repulsion
$V_{C}=2.0$~eV. The two methods give a resonant absorption peak in the XAS for
an incident energy $\omega_{\mathrm{in}}=0.1$~eV. Note that in this comparison we did not use the 
center trick for calculating the RIXS spectra. Instead, Eq.~(3) of the main text has been used. 

\newpage

\section{Supplementary Note~II:~Benchmarks on a multi-orbital corner-shared CuO$_4$ chain} 
Supplementary Figure 2 presents a second comparison of the results obtained
from a multi-orbital Cu$_4$O$_{13}$ cluster, with open boundary conditions. 
The model parameters are the same as those used in the main text.  
Our DMRG approach again gives perfect agreement with the ED result for both the
XAS and RIXS spectra.

\begin{figure}[thbp]
\centering
\vskip -4.5cm
\includegraphics[width=0.7\textwidth]{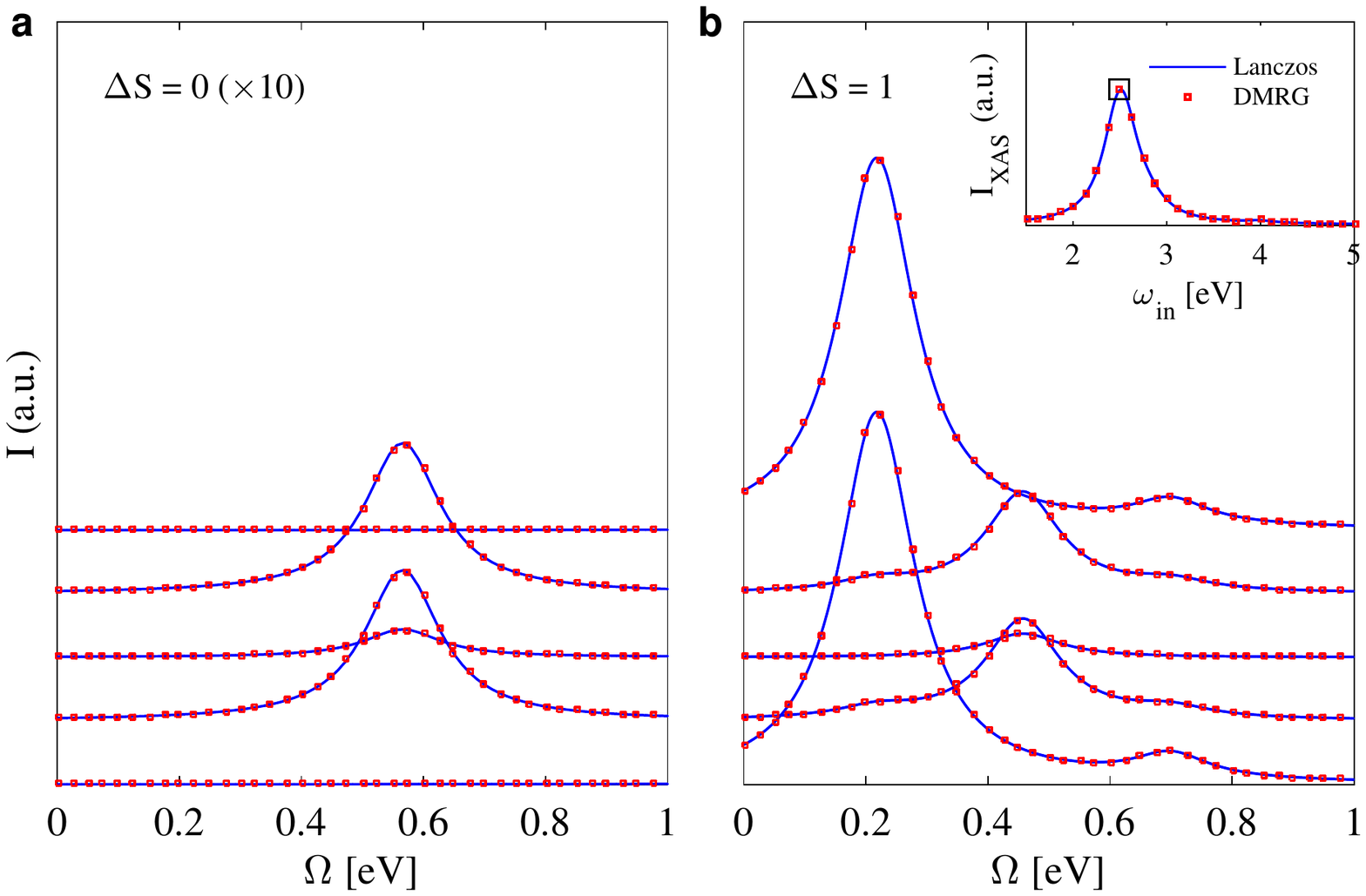}
\vskip -4.5cm
\caption{A comparison of the RIXS response $I(q,\Omega)$ calculated for a 
multi-orbital \emph{pd} model for Sr$_2$CuO$_3$ with 4 unit cells at half filling $N=4$. 
The model parameters are identical to those used in the main text.  
Results are shown for both the (a) $\Delta S = 0$ and (b) $\Delta S = 1$ 
channels, computed at resonance with $\omega_{\mathrm{in}} = 2.5$~eV.  
The spectrum computed with DMRG (red squares) is compared with results obtained from Lanczos ED (solid blue line). 
The inset of panel (b) shows a comparison of the XAS spectrum computed with DMRG [using Eq. (8)] and Lanczos ED. The 
open black box indicates the incident energy $\omega_{\mathrm{in}}$
used to compute the RIXS spectra. Note that in this comparison we did not use the 
center trick for calculating the RIXS spectra. Instead, Eq.~(3) of the main text has been used.  
} \label{supp_fig2}
\end{figure}
\newpage

\section{Supplementary Note~III:~Comparison of the Two Models}
\begin{figure}[h]
\centering
\vskip -1cm
\includegraphics[width=8.5cm]{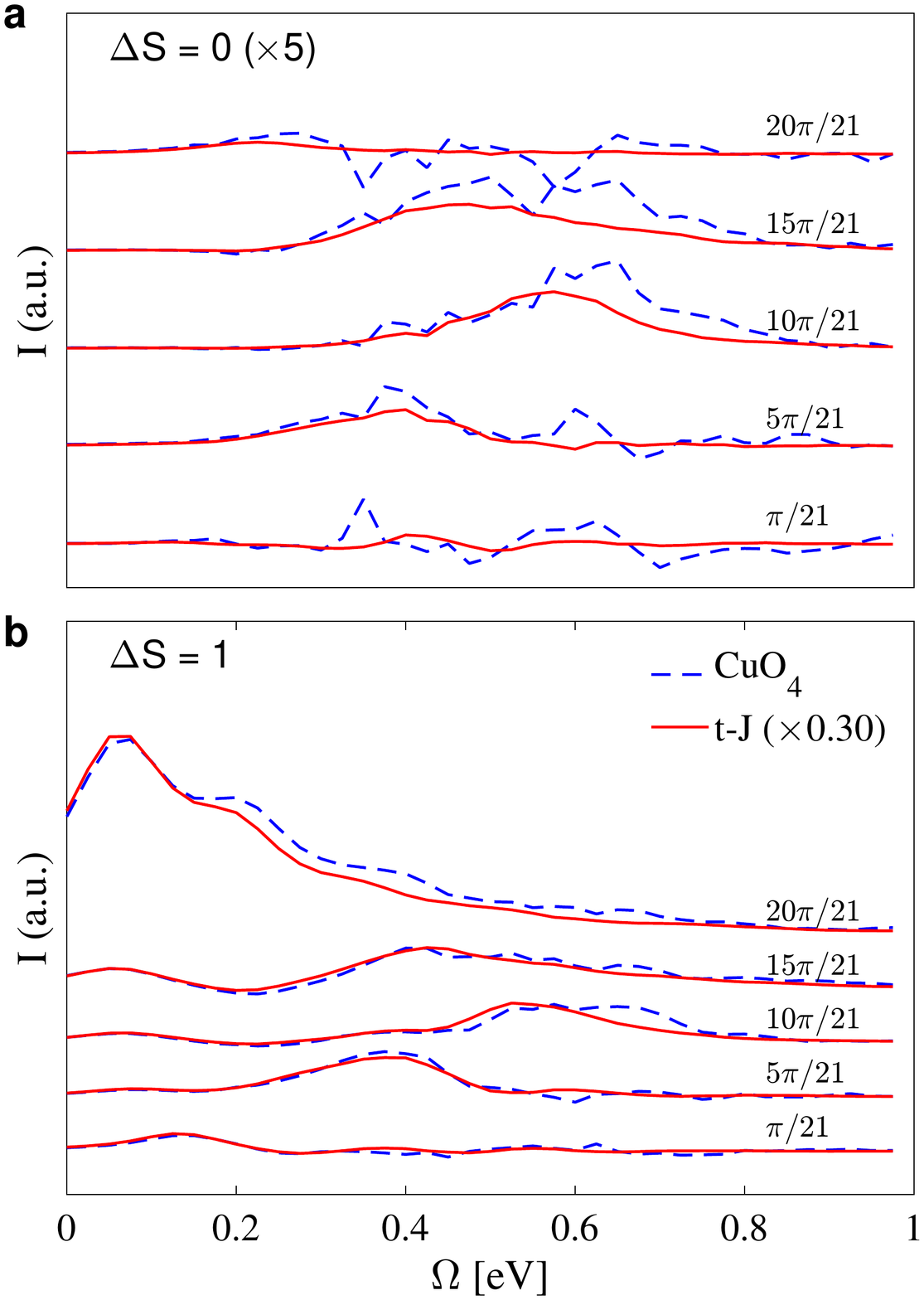}
\vskip -1cm
\caption{
A comparison of the magnetic RIXS excitations computed using DMRG
for a 20-site $t$-$J$ chain (solid red line) and multi-orbital \emph{pd} model (dashed blue line) 
with twenty unit cells at half filling. Results are shown for the (a) $\Delta S=0$
and (b) $\Delta S=1$ channels.
The parameters for the $t$-$J$ model are $t=0.5$~eV, $J=0.325$~eV
and $\omega_{\mathrm{in}}=0.04$~eV.
The parameters for the multi-orbital model are given in the main text. The incident photon energy is $\omega_{\mathrm{in}}=2.5$~eV,  
the inverse core hole lifetime is $\Gamma=0.2$~eV, and the core hole potential is $V_C=2$~eV.
The results for the $t$-$J$ model have been scaled by a factor of $0.30$ such that the maximum
intensity of the $\Delta S = 1$ excitations is the same at the zone boundary.
} \label{figS3}
\end{figure}

In the main text, we compared results for the magnetic RIXS spectra of the $t$-$J$ and 
multi-orbital $pd$ model, each with 20 unit cells. Here, Supplementary Figure \ref{figS3}  
presents a similar comparison but for 
a different value of the core-hole potential used for the $t$-$J$ model.  

\section{Supplementary Note~IV:~Removing the Elastic Line From the DMRG Calculations}
We first rewrite Eq.~(3) of the main text to explicitly indicate the center site $c$ 
\begin{align}
 I(\textbf{q},\Omega)\propto -\mathrm{Im} \Bigg[ \sum\limits_{i=0}^{L-1}\sum\limits_{\substack{\gamma,\gamma^{\prime}\\\sigma,\sigma^{\prime}}} e^{i \textbf{q}\cdot(\textbf{R}_i-\textbf{R}_c)}
 \langle\alpha_{i,\gamma}| \hat{D}_{i,\gamma^{\prime}} \frac{1}{\Omega-\hat{H}+E_g+\mathrm{i}\eta} \hat{D}^{\dagger}_{c,\sigma^{\prime}}|\alpha_{c,\sigma}\rangle\Bigg]. 
\end{align}
The $\Delta S=0$ contribution computed with DMRG is then given by the expression
\begin{align}
I_{\Delta S=0}(\textbf{q},\Omega)\propto -\mathrm{Im} \Bigg[ \sum\limits_{i=0}^{L-1}\sum\limits_{\sigma=\uparrow,\downarrow} e^{i \textbf{q}\cdot(\textbf{R}_i-\textbf{R}_c)}
 \langle\alpha_{i,\sigma}|\hat{D}_{i,\sigma}\hat{P}\frac{1}{\Omega-\hat{H}+E_g+\mathrm{i}\eta} \hat{P}\hat{D}^{\dagger}_{c,\uparrow}|\alpha_{c,\uparrow}\rangle\Bigg],
\end{align}
where 
$\hat{P}=\mathds{1} - |g\rangle\langle g|$ projects out the ground-state contribution. 
The  expectation values $\langle\alpha_{i,\sigma}|\hat{D}_{i,\sigma}|g\rangle$ (and their hermitian conjugates) are 
calculated in Step~3 of the algorithm, and used in Step~4. 
Here, the contribution to the elastic peak of the spectra is removed by 
the subtraction of $\langle\alpha_{i,\sigma}|\hat{D}_{i,\sigma}|g\rangle$.

The $\Delta S=1$ contribution of the RIXS spectrum is given by
\begin{align}
 &I_{\Delta S=1}(\textbf{q},\Omega)\propto -\mathrm{Im} \Bigg[ \sum\limits_{i=0}^{L-1}e^{i \textbf{q}\cdot(\textbf{R}_i-\textbf{R}_c)}
 \langle\alpha_{i,\uparrow}| \hat{D}_{i,\downarrow}\times\frac{1}{\Omega-\hat{H}+E_g+\mathrm{i}\eta} \hat{D}^{\dagger}_{c,\downarrow}|\alpha_{c,\uparrow}\rangle\Bigg].
\end{align}
In this case, the elastic contribution is absent because $[\hat{H},S^{\text{tot}}_z]=0$, thus 
$\langle\alpha_{i,\sigma}|\hat{D}_{i,\bar{\sigma}}|g\rangle= \langle g|\hat{D}^{\dag}_{i,\sigma}[\omega_\mathrm{in}-\hat{H}_{\mathrm{ch},i}+E_g+\mathrm{i}\Gamma]^{-1} \hat{D}_{i,\bar{\sigma}}|g\rangle=0$ for $\bar{\sigma}=-\sigma$.

\section{Supplementary Note~V:~{\sc {\bf DMRG++}}} 
The {\sc DMRG++} computer program was used for the DMRG results. {\sc DMRG++} is available at
\nolinkurl{https://github.com/g1257/dmrgpp} under a free and open source license,
is maintained, and open for community contributions.

\end{document}